\documentstyle[11pt,aaspp4]{article}

\lefthead{Prosser, Kennicutt, Bresolin et al.}
\righthead{Cepheids in NGC1326-A}

\begin{document}

\title{The HST Key Project on the Extragalactic Distance Scale. \\ 
XXII.  The Discovery of Cepheids in NGC 1326-A$^1$}

\singlespace

\author{Charles F. Prosser\altaffilmark{2,3},
Robert C. Kennicutt, Jr.\altaffilmark{4},
Fabio Bresolin\altaffilmark{5},
Abhijit Saha\altaffilmark{2},
Shoko Sakai\altaffilmark{2},
Wendy L. Freedman\altaffilmark{6},
Jeremy R. Mould\altaffilmark{7},
Laura Ferrarese\altaffilmark{8,9},
Holland C. Ford\altaffilmark{10},
Brad K. Gibson\altaffilmark{11},
John A. Graham\altaffilmark{12},
John G. Hoessel\altaffilmark{13},
John P. Huchra\altaffilmark{14},
Shaun M. Hughes\altaffilmark{15},
Garth D. Illingworth\altaffilmark{16},
Daniel D. Kelson\altaffilmark{12},
Lucas Macri\altaffilmark{14},
Barry F. Madore\altaffilmark{6,17},
Nancy A. Silbermann\altaffilmark{18}
Peter B. Stetson\altaffilmark{19}
}

\altaffiltext{1}{Based on observations
with the NASA/ESA {\it{Hubble Space Telescope}}, obtained at the Space 
Telescope Science Institute, which is operated by AURA, Inc. under NASA
Contract No. NAS5-26555.}
\altaffiltext{2}{NOAO, P.O. Box 26732, Tucson, AZ 85726-6732 USA}
\altaffiltext{3}{Deceased, 15 August 1998.  Charles joined the Key Project
team after earning his doctorate at the University of California at
Santa Cruz, and serving as a postdoctoral fellow at the 
Harvard-Smithsonian Center for Astrophysics.  At the time of his
death he had nearly completed the first draft of this paper.}
\altaffiltext{4}{Steward Observatory, The University of Arizona, Tucson, AZ 85721 USA}
\altaffiltext{5}{European Southern Observatory, Karl-Schwarzschild-Str. 2, Garching, Germany}
\altaffiltext{6}{Carnegie Observatories, 813 Santa Barbara Street, Pasadena, CA 91101 USA}
\altaffiltext{7}{Research School of Astronomy \& Astrophysics, Mount Stromlo Observatory, Weston, ACT, Australia 2611} 
\altaffiltext{8}{Department of Astronomy, MS 105-24, California Institute of Technology, Pasadena, CA 91125 USA}
\altaffiltext{9}{Hubble Fellow}
\altaffiltext{10}{Department of Physics \& Astronomy, Johns Hopkins University,
Baltimore, MD 21218 USA}
\altaffiltext{11}{CASA, University of Colorado, Boulder, CO 80309-7178 USA}
\altaffiltext{12}{Department of Terrestrial Magnetism, Carnegie Institution of Washington, 5241 Broad Branch Road, NW, 
Washington, DC 20015, USA}
\altaffiltext{13}{Department of Astronomy, University of Wisconsin, Madison, WI 53706 USA}
\altaffiltext{14}{Harvard Smithsonian Center for Astrophysics, 60 Garden Street,Cambridge, MA 02138 USA}
\altaffiltext{15}{Institute of Astronomy, University of Cambridge, Cambridge CB3 OHA, United Kingdom}
\altaffiltext{16}{Lick Observatory, University of California, Santa Cruz, CA 95064 USA}
\altaffiltext{17}{NASA/IPAC Extragalactic Database and California Institute of Technology,
Pasadena, CA 91125 USA}
\altaffiltext{18}{IPAC, California Institute of Technology, Pasadena, CA 91125 USA}
\altaffiltext{19}{Dominion Astrophysical Observatory, HIA/NRC of Canada, 5083 West Saanich Road, Victoria, BC, V8X 4M6, Canada}
\clearpage

\setcounter{footnote}{0}

\begin{abstract}

  We report on the detection of Cepheids and the first distance measurement 
to the spiral galaxy NGC~1326-A, a member of the Fornax cluster of galaxies.
We have employed data obtained with the Wide Field and Planetary Camera 2 on
board the Hubble Space Telescope.    Over a 49 day interval, a total of twelve 
$V$-band (F555W) and eight $I$-band (F814W) epochs of observation were obtained.
Two photometric reduction packages, ALLFRAME and DoPHOT, have been employed to
obtain photometry measures from the three Wide Field CCDs.  Variability 
analysis yields a total of 17 Cepheids in common with both photometry
datasets, with periods ranging between 10 and 50 days.  Of these 14 Cepheids
with high-quality lightcurves are used to fit the $V$ and $I$ period-luminosity 
relations and derive apparent distance 
moduli, assuming a Large Magellanic Cloud distance modulus
$\mu_{\rm LMC} = 18.50 \pm 0.10$ mag and color excess
E(B$-$V) $=$ 0.10 mag.     Assuming A(V)/E(V$-$I) $=$ 2.45, 
the DoPHOT data yield a true distance modulus to NGC~1326-A of $\mu_{\rm o} = 
31.36 \pm 0.17$
(random) $\pm 0.13$ (systematic) mag, corresponding to a distance of
$18.7 \pm 1.5$ (random) $\pm 1.2$ (systematic) Mpc.
The derived distance to NGC~1326-A is in good agreement with
the distance derived previously to NGC~1365, 
another spiral galaxy member of the Fornax cluster.  However
the distances to both galaxies are significantly lower than to
NGC~1425, a third Cepheid calibrator in the outer parts of the cluster.

\end{abstract}


\clearpage

\section{INTRODUCTION}

This paper is one in a series of the {\it Hubble Space Telescope} Key
Project on the Extragalactic Distance Scale; a recent overview of this 
project may
be found in Kennicutt, Freedman, \& Mould (1995). The primary goal is to obtain
a measure for the Hubble constant (H$_0$) within an accuracy of $\pm$10\%, by
employing Cepheid-derived distances for nearer galaxies (redshifts generally 
within $\sim$1500 km s$^{-1}$)  to calibrate several secondary distance
indicators (Tully-Fisher relation, fundamental plane, 
surface brightness fluctuation method, supernovae distance indicators).
In this paper we present the results from a Cepheid monitoring program 
of NGC~1326-A, a spiral galaxy belonging to the Fornax cluster of galaxies.  

NGC~1326-A ($\alpha_{\rm 1950} = 3^{\rm h} 23^{\rm m}$,
$\delta_{\rm 1950} = -36^\circ 31'$) is the third galaxy in the Fornax cluster
region which has been studied in this series.  Observations of NGC~1365
and NGC~1425 have been reported by Madore et al. (1998, 1999), 
Silbermann et al. (1999) and Mould et al.
(1999), respectively.  A survey of the Fornax cluster
has been reported by Ferguson (1989).    Fornax is a relatively compact 
cluster on the sky (core radius $\sim$ 0.7 deg), containing 350 identified 
members, of which most are elliptical galaxies.  The heliocentric radial 
velocity of the Fornax Cluster is 1441 km s$^{-1}$, with a dispersion of 
342 km s$^{-1}$
(Madore et al. 1999).   NGC~1326-A lies at a projected separation of 
2.0 deg from the cluster center, with a 
measured heliocentric velocity of 1836 km s$^{-1}$ 
(da Costa et al. 1991).  The galaxy is classifed as type SBm pec
by de Vaucouleurs et al. (1991).  It lies $\sim$2.5 arcmin away from another
galaxy (NGC 1326-B), and may be interacting with that system, although
their radial velocities differ by over 800 km s$^{-1}$, which may 
alternatively indicate that they are a line-of-sight pair.

This paper is organized as follows.
A description of the observations and data reduction is 
provided in \S2, a description of the photometric reductions is given in \S3,   
and the identification of variable stars given in \S4.  In \S5, the Cepheid 
sample for NGC 1326-A is reviewed,  and in \S6 the resulting period-luminosity
relations and distance modulus are presented,  with further discussion and
conclusions presented in \S7.   A series of appendices provide additional
photometry, light curves, and a summary of other variable stars found.

\section{OBSERVATIONS AND DATA REDUCTION}

  NGC~1326-A was observed using the Hubble Space Telescope Wide Field and 
Planetary
Camera 2 (HST/WFPC2; Biretta et al. 1996) over a 49 day interval during 
1997 August/September.
In Figure 1, we show the WFPC2 field of
view of the current observations superimposed on a V-band CCD image of 
NGC~1326-A, obtained with the 2.5-m telescope at the Las Campanas Observatories.

In Table 1 we list the WFPC2 observations for NGC 1326-A, giving the image 
identification
code, UT Date of observation, the corresponding heliocentric Julian Date 
(mid-exposure), filter, number of individual exposures in the epoch, and 
combined total exposure time.  Each epoch consisted of 3 or 4 exposures taken
over two orbits, to provide for better cosmic ray discrimination. 
Twelve epochs in $V$ (F555W) and eight epochs in $I$ (F814W) were obtained.
As in earlier studies, the epochs roughly follow a power-law sampling 
distribution to reduce aliasing effects and maximize phase coverage over 
the $\sim$10--60 day   
Cepheid period range (Freedman et al. 1994, Madore \& Freedman 1991). 
All observations were preprocessed through the standard reduction pipeline at
STScI (Holtzmann et al. 1995a), which includes bias level 
subtraction, superbias and dark frames subtraction, and flatfielding,
using the most recent calibration files available.  
Our own post-pipeline processing 
included the masking of bad columns/pixels, masking of the vignetted edges 
of the CCDs, and 
multiplication by a pixel area correction map to correct for geometric
distortion in WFPC2 images (Hill et al. 1998).  Finally, the calibrated images
were converted to integer format for use in our photometric reduction software, 
after being multiplied by a factor of four in order to preserve data precision. 

\section{PHOTOMETRIC REDUCTIONS}

    As in previous studies in this series, photometric analysis of the WFPC2
images was performed independently using the software routines DAOPHOT
II/ALLFRAME (Stetson 1994), and a variant of DoPHOT (Schechter et al. 1993),
as described in Saha et al. (1994).  In this section, we
briefly summarize the separate reductions and then compare measures using
a selected list of `secondary standard stars' from the WFPC2 fields.   

Preliminary reductions revealed very few stars on the PC chip with
reliable photometry, and a full DoPHOT reduction did not yield any
usable Cepheid candidates in the PC dataset.  As a result, full 
photometric reductions 
in both software routines and subsequent analysis for variables were 
only carried out for the Wide Field CCDs WF2-WF4.

\subsection{ALLFRAME Photometry}

For the ALLFRAME reduction, no cosmic-ray corrections were
applied to the individual galaxy frames.  A master star list was
created by first making two median-filtered images, one from the 40 cosmic 
ray split F555W images and another from the 24 F814W images.  
These were later summed to generate one clean frame.
We adopted this method so as to include both extremely red and blue
stars in the master star list.  The star list was created 
with the automated star-finding routines from the DAOPHOT and 
ALLSTAR packages.  This list was fed into ALLFRAME and used to extract
profile-fitting stellar photometry from the 64 individual frames.

The procedures used to convert F555W and F814W instrumental magnitudes to the 
calibrated Landolt (1992) system are described in detail in Hill et al. (1998). 
For completeness, however, a  brief summary follows here.  The instrumental
magnitudes were first brought to the 0\arcsec.5 aperture system of
Holtzman et al. (1995).  This was done by identifying 
30 to 40 isolated bright stars on each chip in the median-filtered master
image, and performing concentric
aperture photometry on each star, using 12 apertures ranging in radius
from 0\arcsec.15 to 0\arcsec.50.  All other stars were subtracted from the
original image, leaving a clean image adequate for good sky estimates.
Using a growth--curve analysis provided by DAOGROW (Stetson 1990), 
the 0\arcsec.5 aperture magnitude for each star was finally determined, 
and then compared with the corresponding PSF magnitude to calculate the 
aperture correction.  For each filter and chip combination, the aperture 
correction was determined by taking the weighted mean of the aperture 
corrections of all the selected isolated stars.  
The typical star-to-star variation in aperture corrections determined
for single stars is $\pm$0.05 mag rms, so the mean corrections are determined
to an internal precision of roughly 0.01--0.02 mag, as listed in Table 2.

The transformation equation to convert the 0\arcsec.5 aperture magnitudes, $m$,
to the Hill et al. (1998) system, $M$, is expressed as:

\begin{equation}
M = m + 2.5\log t + C1 + C2 * (V-I) + C3 * (V-I)^2, 
\end{equation}

\noindent where $t$ is the exposure time and C1, C2, and C3 are constants.
The coefficients C2 and C3 for the color--corrections were adopted
from Holtzman et al. (1995); they are the same for all four chips.
The constant C1 consists of several terms: the long--exposure
WFPC2 magnitude zero points, the ALLFRAME magnitude zero point,
a correction for multiplying the original image by four before
converting them to integers, a gain ratio term due to the difference
between the gain settings used for NGC~1326-A and for the Holtzman et al.
data (7 and 14 respectively), and a correction for the pixel area
map which was normalized differently from Holtzman et al. (1995).
Each term is discussed in detail by Hill et al. (1998).
Values of the C1 terms are listed in Table 2.  As was the case for
other papers in this series, we have adopted the ``long exposure"
zeropoints of Hill et al. (1998) for both the ALLFRAME and DoPHOT photometry.

Recently Stetson (1998) has reanalyzed the zeropoint calibration of 
WFPC2 taking into account possible second-order charge transfer effects.
In order to maintain consistency with previous papers in this series
we have not adopted the corrections derived in Stetson (1998), adopting
instead the Hill et al. (1998) zeropoints with the 0.05 mag correction
for charge transfer effects.  Any differences between the two scales
should be very small in comparison to the other uncertainties in the
Cepheid distance to NGC~1326-A.  A new analysis of the photometric
zeropoint is currently being carried out by P. Stetson, and will be
reported in a future paper.

\subsection{DoPHOT Photometry}

    The DoPHOT reduction followed the procedures desribed in Ferrarese
et al. (1996, 1998).  Additional
detailed discussion of the procedure may also be found in Saha et al. (1996).
We briefly summarize the reduction procedure here, and mention aspects specific 
to the current study of NGC~1326-A.

Prior to running DoPHOT, the 3-4 individual exposures at each epoch in Table 1
were coadded into a single image, with cosmic ray events identified and 
rejected in the process.  After coalignment of the images 
from each epoch, master F555W and F814W frames were derived from median
filtering of the combined epoch observations.   DoPHOT was then run on
these deep images to produce a master object list, which was then 
transformed and applied as input when running DoPHOT on each separate epoch.   
The output PSF photometry (m$_{\rm psf}$) from DoPHOT is then scaled by the
exposure time (t) and calibrated onto
the Holtzmann et al. (1995) 0\arcsec.5 `ground system' (m$_{\rm grd}$)
through the use of aperture correction (AC) and zero point (ZP) terms:

\begin{equation}
{\rm m}_{\rm grd} = {\rm m}_{\rm psf} + 30.0 + 2.5log(t) + {\rm AC} + ZP
\end{equation}

A different approach to calculating the aperture corrections is used
in the DoPHOT reductions (Saha et al. 1996).  The positional variation in
the corrections is determined from external data with a clean background.
Due to the paucity of bright isolated stars in NGC~1326-A, the mean
aperture corrections were derived from single-epoch observations of the 
Local group dwarf spheroidal galaxy Leo I, as described in Saha et al. (1996).
This procedure has been tested extensively for other galaxies in the
project and has been shown to reproduce the aperture corrections within
the uncertainties over a wide range of WFPC2 data.  The coefficients 
used to derive these 
AC values are given in Table 3, along with the zero point (ZP) values used.
The ZP values include corrections for the gain ratio term among the different
chips.  The rms scatter in AC measurements for single stars, after fitting
of the spatial variation, is $\pm$0.035 mag rms, and the rms uncertainties
in the average corrections are roughly an order of magnitude smaller.

The mean F555W and F814W magnitudes of the Cepheids and the secondary
standard stars were converted to standard $V$ and $I$ magnitudes, using
the color correction coefficients
from Holtzmann et al. (1995, Table 3): 

\begin{equation}
{\rm V} = {\rm F555W} - 0.045({\rm F555W} - {\rm F814W}) 
+ 0.027({\rm F555W} - {\rm F814W})^2,
\end{equation}
\begin{equation}
{\rm I} = {\rm F555W} - 0.067({\rm F555W} - {\rm F814W})
+ 0.025({\rm F555W} - {\rm F814W})^2.
\end{equation}

%
%
%

\subsection{Comparison of DoPHOT and ALLFRAME Photometry}

To compare the DoPHOT and ALLFRAME magnitudes, we selected
stars in each the 3 WF chips which were
relatively isolated, showed low formal errors in their magnitude measures, 
were distributed over the CCD,  and covered a magnitude range in common with
the detected Cepheid variables.
The locations and DoPHOT photometry of these `secondary standard' stars
are given in Appendix A.   Figure 2 shows plots of the $V$ and $I$ magnitude 
differences (DoPHOT $-$ ALLFRAME) for the WF2-WF4 chips.
The corresponding mean differences and errors are given in Table 4.  The 
magnitudes of the secondary standard stars show reasonable consistency,
especially when taking into account the relatively large distance and
high crowding level in this galaxy.  The Cepheids themselves show somewhat
larger differences, reflecting the fact that the Fornax cluster lies
near the limit at which accurate crowded-field photometry can be obtained
with WFPC2.

\section{IDENTIFICATION OF VARIABLE STARS}

In the case of the ALLFRAME data set the search for variables was
carried out with a variation of the correlated variability test of
Welch \& Stetson (1993), as implemented in the algorithm for the
automated detection of Cepheids described by Stetson (1996). This
method, which relies on the temporal correlation of magnitude
residuals of real variables, takes advantage of the CR split nature of
the HST observations.

Variability analysis of the DoPHOT data set 
was performed on the F555W observations, following the
general technique detailed in Saha \& Hoessel (1990).  In order to be 
flagged as a variable, a star had to be measured in at least 8 out of 
the 12 $V$-band epochs.  Stars with nearby neighbors 
(within 2 pixels) that contributed more than 50\% of the total light 
were rejected.   Variable stars were identified using a simple reduced
$\chi^2$ criterion, with the requirement that the variability be significant
at the $>$99\%\ confidence level, or $\chi^2 > 8$ (Ferrarese et al. 1998).
Stars with confidence levels above 99\%\ but with lower $\chi^2$ values 
were tested further for periodicity, using a variant of the 
Lafler-Kinman (1965) method of phase dispersion minimization.  
Stars with $\Lambda \ge 2.8$  were flagged as candidate variables, 
where $\Lambda$ is as defined in Lafler \& Kinman (1965), and implemented 
in Saha \& Hoessel (1990). 

NGC~1326-A is one of the most distant galaxies in the Key Project, and
this combined with its relatively low luminosity yields a smaller potential
sample of candidate variables than most of the other galaxies in the project.
For this study, we performed an additional check on the significance
of the variability, by performing random redistribution of the observational
dataset to check directly the random probability of detecting each candidate
as a variable.  We performed 100 random resamplings for each candidate,
and applied the reduced $\chi^2$ and $\Lambda$ tests described above to these
data.  This `bootstrapping' technique allowed us to eliminate numerous  
spurious variables which passed the variability tests due to a single outlying 
observation, usually from a bad pixel or cosmic ray event.  A final check 
for spurious variables was made by visually inspecting and blinking 
the images for each candidate.

Periods for the candidate variable stars were determined by phasing the data 
for all periods between 3 and 100 days, and applying the Lafler-Kinman
phase dispersion minimization criterion.  Because of the logarithmic
sampling of the data, period aliases are not a serious problem, and
in the rare ambigous cases the data were checked for spurious observations
and the variability analysis was performed interactively.  Serious aliasing
was often a problem for stars with periods longer than our observing window
(49 days), so only those Cepheids with $P < 50$ days are analyzed in this
paper.

\section{CEPHEID VARIABLE STARS}

\subsection{Cepheid Sample} 

The final list of Cepheid variable stars in NGC~1326-A is composed of 
stars which meet the variability criteria in both of the ALLFRAME and 
DoPHOT data sets.  When matching the candidate lists we also performed several
other checks on the properties of the stars.  Stars with lightcurve shapes
that clearly differ from Cepheids, stars with low variability amplitude 
or flat-bottomed lightcurves, those with highly abnormal colors, or
those with strong variability in $V$ but little or none in $I$ were all
rejected from the final list.  These tests were aimed at rejecting 
non-Cepheid variable stars and Cepheids with unresolved companions.  
Stars located in crowded regions were also rejected, according to the
criteria given in \S 4.  None of these rejected candidates were included
in the final Cepheid sample, but a list of indeterminate variables and
long-period variables is given in Appendix B.

Adopting these criteria yielded 17 high-quality Cepheids in common between
the DoPHOT and ALLFRAME lists.  These stars, along with their astrometric
positions are listed in Table 5, in order of decreasing period.
The absolute positions are accurate to within the pointing accuracy of
HST ($\sim$1\arcsec), while the relative positions should be accurate
to approximately 0.\arcsec2.  
Finder charts for the Cepheids (as well
as the other variable stars) are given in Figure 3, and enlarged images
of the regions surrounding each Cepheid are given in Figure 4.
DoPHOT and ALLFRAME mean photometry is given in Tables 6
and 7, respectively.  These tables include two candidates, C8 and C10, which
yielded high-quality lightcurves with DoPHOT only.  Such inconsistency
between the two programs is usually caused by semi-resolved companions
to the Cepheids, so as a precaution these are listed but
are not included in the fitting of the period-luminosity relation.  
Individual DoPHOT magnitudes in F555W and F814W  (and ALLFRAME $V$ and
$I$ magnitudes) at each epoch are 
provided in Appendix C\footnote{Additionally, these values may be 
obtained through the HST Key Project Archives
at http://www.ipac.caltech.edu/H0kp.}.  Finally, the phased lightcurves
of the Cepheids in F555W and F814W are shown in Figure 5.

\subsection{Mean Magnitudes}

Tables 6 and 7 list both intensity averaged 
(`av' subscript) and  phase weighted (`ph' subscript) mean magnitudes
for each Cepheid, defined as 

\begin{equation}
{\rm m}_{\rm av} = -2.5{\rm log}_{10}\sum_{i=1}^n {1\over{n}}10^{-0.4m_i},
\end{equation}
\begin{equation}
{\rm m}_{\rm ph} = -2.5{\rm log}_{10}\sum_{i=1}^n 0.5(\phi_{i+1} - \phi_{i-1})10^{-0.4m_i},
\end{equation}

\noindent
where $n$ is the total number of observations and $m_i$ and $\phi_i$ are the
magnitude and phase of the $i$th observation in order of increasing phase.
The variability levels of reduced $\chi^2$ for DoPHOT and $\sigma$ for ALLFRAME
are provided in the last columns of the tables.
In the case of the F814W ($I$) observations,  due to the fewer phase points
available the phase weighted means have been
adjusted by the small correction factor based on the $V$ to $I$ amplitude 
ratio, following the procedure described in Freedman et al. (1994).  As may be 
expected, differences between ${\rm m}_{\rm av}$ and ${\rm m}_{\rm ph}$
are larger when the phase coverage is less uniform.

The periods determined from the DoPHOT and ALLFRAME fitting show very
good agreement, usually within 5\%\ or better.  The mean uncertainty
in the intensity and phase-weighted mean magnitudes is $\pm$0.1 mag,
which is comparable to the consistency of the DoPHOT and ALLFRAME
magnitudes, as summarized in Table 4.  
This level of agreement in magnitude scales is lower than in most
of the other galaxies in the Key Project sample, but is comparable
to the errors found 
previously in galaxies near the distance limit of this project, e.g., 
NGC~4414 (Turner et al. 1998), NGC~1365 (Silbermann et al. 1999), and
NGC~4548 (Graham et al. 1999).  We have included these 
uncertainties in the total error budget for the distance to NGC~1326-A
(\S 6).

Figure 6 shows the $V$ vs. $V - I$ color magnitude diagram for 
NGC 1326-A, based on the DoPHOT photometry of the three WF chips. 
Phase-averaged mean magnitudes for the Cepheids are overplotted,
along with the expected $P = 10 - 60$ day locus line for zero reddening and 
an assumed distance modulus of 31.4 mag.

\section{PL RELATIONS AND THE DISTANCE TO NGC 1326-A}

\subsection{PL Relations and Apparent Distance Moduli}

Apparent distance moduli to NGC~1326-A in $V$ and $I$ were determined
using the final set of Cepheids listed in Tables 5--7, and the 
calibrating PL relations from Madore \& Freedman (1991):

\begin{equation}
M_V  = -2.76(\log P - 1.0) - 4.16  
\end{equation}

\begin{equation}
M_I = -3.06(\log P - 1.0) - 4.87
\end{equation}

\noindent
These calibrations are based on a sample of 32 Cepheids in the LMC,
and assume an LMC true modulus and average line-of-sight reddening  of
$18.50 \pm 0.10$ and  $E(B-V) = 0.10$ mag respectively. 
As in all other papers in this series, the slopes of the PL relations 
are fixed to the values given in equations (5) and (6), and only the
mean shift from the LMC relation is fitted. This minimizes any biases 
which might be introduced by 
incompleteness at short periods in the NGC~1326-A sample.  

The PL relations for the NGC~1326-A Cepheids from the DoPHOT data
are shown in Figure 7.
The relations were fitted using 14 of the Cepheids in Tables 5--7, 
shown as solid points in Figure 7.
Stars C08 and C10 were not used in the fit because they were identified
cleanly only in the DoPHOT data set.  C11 and C19 were not fitted because they
show very anomalous colors in the DoPHOT data, differing by more than
half a magnitude from the expected colors for Cepheids of their respective
periods, compared to an rms photometric error of $\sim$0.15 mag in $V-I$ 
(C19 also has a period below
10 days).  Such anomalous colors are often a sign of an unresolved
blend with a companion star, so as a conservative measure C11 and C19
were excluded from the PL fitting.  C04 was also rejected because
it lies more than 3$\sigma$ above the mean PL relation, probably due 
to an unresolved blend with another star, which would also account
for its low amplitude.  The Cepheids which were
not included in the fits are shown as open circles in Figure 7.
The 14 stars used to
fit the PL relation have periods in the range 10--50 days.

The fits to the $V$ and $I$ PL relations were performed as
described in Freedman et al. (1994a).   Fitting the DoPHOT periods and
phase weighted magnitudes yields apparent moduli $\mu_V = 31.36 \pm 0.06$ mag,
$\mu_I = 31.36 \pm 0.06$ mag, and $E(V-I) = 0.00 \pm 0.05$ mag, where the 
quoted uncertainties refer to the rms dispersion in the mean PL fits only.
The corresponding moduli
from the ALLFRAME data set are $\mu_V = 31.39 \pm 0.07$ mag, $\mu_I = 31.34 \pm
0.06$ mag, and $E(V-I) = 0.05 \pm 0.03$ mag.
The best fitting PL relations are
shown by the solid lines in Figure 7. The dashed lines, drawn at
$\pm0.54$ mag for the $V$ PL plot, and $\pm0.36$ mag for the $I$ PL
plot, represent the 2$\sigma$ scatter of the best fitting PL relation for 
the LMC Cepheids (Madore \& Freedman 1991). 

\subsection{Extinction and True Distance Modulus}

The true distance modulus to NGC~1326-A, $\mu_0$, is given by 

\begin{equation}
\mu_0 = \mu_V - A(V) = \mu_I - A(I)
\end{equation}

\noindent 
where the $V$ and $I$ band absorption coefficients $A(V)$ and
$A(I)$ obey the relation $A(V)/E(V-I) = 2.45$. This is consistent with
the extinction laws Dean, Warren, \& Cousins (1978), Cardelli, Clayton \& 
Mathis  (1989), and Stanek (1996) and assumes $R_V = 
A(V)/E(B-V) = 3.3$.  Applying this to the DoPHOT PL fits given above
yields a true distance modulus $\mu_0 = 31.36 \pm 0.11$ mag, again with
the quoted uncertainty only reflecting the formal dispersion in the
individual Cepheid distance moduli.  The ALLFRAME data give a similar
value for the true modulus, $\mu_0 = 31.27 \pm 0.08$ mag.
The corresponding distances for the ALLFRAME and DoPHOT
fits are 18.7 Mpc and 17.9 Mpc, respectively.  The uncertainties 
associated with these distance are discussed below.

\subsection{Error Budget}

The error budget in the determination of the true distance modulus is
summarized in Table 8.  To maintain consistency with the previous papers
in this series we closely follow the analysis of Phelps et al. (1998).
This paper contains a detailed discussion and justification for the
individual error terms, and here we only summarize those aspects 
which are specific to NGC~1326-A.  

The first error terms in Table 8 ({\it a-c, S1}) 
are those arising from the LMC-based
calibration of the PL relation, including errors in the assumed LMC
distance and uncertainties in the calibrating Cepheid photometry.
Because this uncertainty affects all Key Project galaxies exactly the
same way, it is of a systematic nature.  The second set of terms 
({\it d-e, R1}) include
uncertainties in the LMC photometric zeropoints, as discussed in detail
by Hill et al. (1998) and Stetson et al. (1998).  These introduce a
significant uncertainty in the true distance modulus because of the
strong propagation of color errors in the dereddened modulus.  
Uncertainties in the PL fits and reddening to NGC~1326-A introduce
an additional random error ({\it f-g, R2}).  

Another source of systematic uncertainty in the distance moduli 
is the uncertain dependence of the Cepheid PL relation on metal
abundance (Kennicutt et al. 1998).  However this is not a serious
problem for NGC~1326-A.  Spectrophotometry of five HII regions in
the galaxy by Kennicutt et al. (1999) yields a mean oxygen abundance
$12 + \log(O/H) = 8.48 \pm 0.16$ when calibrated on the scale of
Zaritsky, Kennicutt, \& Huchra (1994).  This is virtually identical
to the mean value for the LMC of 8.50 $\pm$ 0.1 when calibrated on the same
system.  Consequently any metallicity effect should be minimal,
and the error term ({\it S2}) in Table 8 only reflects the 
effect of the uncertainty in the metal abundance itself, when
propagated with the Cepheid metallicity dependence in Kennicutt et al. (1998).

Combining these error terms in quadrature with the PL fits from the
DoPHOT photometry yields final values for the distance modulus of
NGC~1326-A of $\mu_0 = 31.36 \pm 0.17$ (random) $\pm 0.13$ (systematic) mag.
The corresponding distance is 18.7 $\pm$ 1.5 (random) $\pm$ 1.2 (systematic)
Mpc.  The ALLFRAME analysis yields a distance that is 3.3\%\ lower
($\mu_0$ = 31.27, $d$ = 17.9 Mpc), with the same uncertainties.
In this case we preferentially quote the DoPHOT distance, because
the PL relations are somewhat tighter and the fitted distance modulus
is less sensitive to whether the shortest-period variables are 
excluded.  However both sets of distances are consistent within
their quoted errors.

\section{DISCUSSION}

NGC~1326-A is a relatively unstudied galaxy, and no previous distance
measurements could be found in the literature.  An extensive literature
exists on the distance of the Fornax cluster, of course, as discussed
in Madore et al. (1999).  The distance of 18.7 Mpc that we derive to 
NGC~1326-A, although
uncertain at the the 10\%\ level, is in excellent agreement with
the distance of 18.2 $\pm$ 1.7 (random) $\pm$ 1.5 (systematic) Mpc 
derived to another Fornax
cluster member, NGC~1365, based on measurements of 52 Cepheids 
(Silbermann et al. 1999).  However the distances we derive to NGC~1326-A
and NGC~1365 are significantly smaller than that derived for the 
third Fornax cluster galaxy in our sample, NGC~1425, at 22.2 $\pm$ 1.7 (random)
$\pm$ 1.8 (systematic) Mpc (Mould et al. 1999).  A detailed comparison of these
distances and their interpretation is given in Mould et al. (1999).

\acknowledgments

We acknowledge the assistance of Matthew Trewhella in retrieving from 
archive the CCD image used in Figure 1.  We also thank the anonymous
referee for comments which improved the paper significantly.
We gratefully acknowledge the support of the NASA and STScI support staff.
Support for this work was provided by NASA through grant G0-2227-87A from STScI.

\appendix
\section{SECONDARY STANDARDS IN THE NGC~1326-A FIELD}

Tables 9--11 list photometry for the stars selected as secondary standard
stars in each of the WF chips, as described in \S 3.3 and used
in the comparison shown in Figure 2.   The tables list for each star
the DoPHOT (x,y) coordinates on the CCD chips, 
DoPHOT F555W and F814W ground system magnitudes,  the corresponding
calibrated $V$ and $I$ DoPHOT magnitudes, and the calibrated $V$ and $I$ 
ALLFRAME magnitudes.

\section{OTHER VARIABLE STARS IN NGC~1326-A}

Tables 12--13 list identification information and photometry for 17
miscellaneous variable stars in NGC~1326-A.  These stars include Cepheids 
which were excluded from the final sample because of uncertain periods,
large magnitude errors, or inconsistencies between the ALLFRAME and DoPHOT
data sets, long-period variables with periods longer than our sampling
window, or other types of variable stars, as described in \S 4.2.  
Table 12 contains positional information and notes which provide
additional information.  Table 13 lists DoPHOT photometry and period
information (when available) for these stars.

\section{INDIVIDUAL EPOCH PHOTOMETRY OF CEPHEIDS}

Tables 14--15 list the DoPHOT F555W and F814W (ground system) photometry
for the final sample of Cepheids on an epoch-by-epoch basis.  Tables 16--17
list the corresponding ALLFRAME photometry.

\clearpage

\centerline{\bf Figure Captions}
\medskip

{Fig. 1.--- V-band ground based image of NGC 1326-A, with the 
approximate dimensions
of the HST WFPC2 field of view overplotted.  The bright galaxy on the 
edge of the image is NGC~1326-B.}  North is to the top and east to the left.

{Fig. 2.--- Comparison between the DoPHOT/ALLFRAME V and I magnitudes 
among the secondary standard stars from the three WF CCDs.}

{Fig. 3.--- Median-filtered V-band images of the four WFPC2 chips.   
The locations of the Cepheids and other variable stars recovered in this 
study are indicated by circles, and are  
labeled as listed in Tables 5 and 12 . In each frame, the pyramid for the WFPC2
camera is in the lower lefthand corner.}

{Fig. 4.--- Expanded finder charts for the Cepheids (Table 5) and the other
variable stars (Table 12).  Each panel is 50 $\times$ 50 pixels on the WFC, 
corresponding to 5\arcsec\ square on the sky.}

{Fig. 5.--- Light curves for the Cepheid variables, based on the
DoPHOT photometry. The $V$ magnitudes are plotted as solid points, the
$I$ magnitudes as open squares. The errorbars reported by DoPHOT are
indicated.}


{Fig. 6.---  $V$ vs. $V-I$ color magnitude diagram for NGC 1326-A, based on
DoPHOT photometry from median filtered images of the four WFPC2 chips.  
The recovered Cepheid sample is shown by large points. 
The expected ridge line for Cepheids with $10 \le P \le 60$ days and 
zero reddening is drawn for reference, assuming an apparent distance modulus 
of 31.4 mag.  Photometric error bars for a typical single epoch of observation 
are also shown.}

{Fig. 7.--- PL relations in $V$ (top) and $I$ (bottom) for Cepheids in 
NGC~1326-A, based on the DoPHOT photometry.  Cepheids included in the
fitting are shown as solid points, while open circles denote Cepheids
that were rejected from the fit, as described in \S 6.1.  The solid
line shows the fitted PL relation, while the dashed lines show the 
2$\sigma$ intrinsic width of the PL relation in the LMC.}

\newpage

\begin{deluxetable}{cccccc}
\tablenum{1}
\tablecolumns{6}
\tablewidth{0pc}
\scriptsize
\tablecaption{NGC 1326-A Journal of Observations}
\tablehead{
\colhead{STScI} &
\colhead{1997} &
\colhead{Average} &
\colhead{} &
\colhead{No.} &
\colhead{Exposure}  \nl
\colhead{Rootname ID} &
\colhead{UT Date} &
\colhead{HJD} &
\colhead{Filter} &
\colhead{Exp.} &
\colhead{Time (sec)}
} 
\startdata
U34L2601M  & AUG06   & 2450667.0952   & F555W  & 3  &  3400 \nl
U34L2701M  & AUG09   & 2450670.1196   & F555W  & 3  &  3400 \nl
U34L2801M  & AUG12   & 2450673.0080   & F555W  & 3  &  3900 \nl
U34L2901M  & AUG15   & 2450675.9062   & F555W  & 4  &  5200 \nl
U34L3001M  & AUG18   & 2450679.2000   & F555W  & 4  &  5200 \nl
U34L3101R  & AUG23   & 2450684.3708   & F555W  & 3  &  3900 \nl
U34L3201R  & AUG27   & 2450687.7434   & F555W  & 4  &  5200 \nl
U34L3301R  & SEP01   & 2450693.3174   & F555W  & 3  &  3900 \nl
U34L3401O  & SEP05   & 2450697.2925   & F555W  & 4  &  4600 \nl
U34L3501R  & SEP12   & 2450704.2744   & F555W  & 3  &  3400 \nl
U34L3601R  & SEP17   & 2450709.3821   & F555W  & 3  &  3400 \nl
U34L3701R  & SEP24   & 2450716.1699   & F555W  & 3  &  3400 \nl
  ...      &  ...    &     ...        &  ...   &... &  ...  \nl
U34L2604M  & AUG06   & 2450667.1765   & F814W  & 3  &  3600 \nl
U34L2704M  & AUG09   & 2450670.2008   & F814W  & 3  &  3600 \nl
U34L2804M  & AUG12   & 2450673.0917   & F814W  & 3  &  3900 \nl
U34L3104R  & AUG23   & 2450684.4542   & F814W  & 3  &  3900 \nl
U34L3304R  & SEP01   & 2450693.4007   & F814W  & 3  &  3900 \nl
U34L3504R  & SEP12   & 2450704.3557   & F814W  & 3  &  3600 \nl
U34L3604O  & SEP17   & 2450709.4633   & F814W  & 3  &  3600 \nl
U34L3704R  & SEP24   & 2450716.2515   & F814W  & 3  &  3600 \nl
\enddata
\end{deluxetable}

\begin{deluxetable}{cccccc} 
\tablenum{2}
\tablecolumns{6}
\tablewidth{0pc}
\scriptsize
\tablecaption{Coefficients Used for ALLFRAME Photometry}
\tablehead{
\colhead{Chip} & 
\colhead{Filter} & 
\colhead{Stars} &
\colhead{Aperture Correction} &
\colhead{WFPC2} &
\colhead{C1} \nl
\colhead{} & 
\colhead{} & 
\colhead{} &
\colhead{Zeropoint\tablenotemark{a}} &
\colhead{Coeff.\tablenotemark{b}} 
}
\startdata
WF2  & F555W & 51 & $-$0.027$\pm$0.017 & 22.522 & $-$0.984  \nl
WF2  & F814W & 51 & $+$0.006$\pm$0.006 & 21.616 & $-$1.857  \nl
WF3  & F555W & 22 & $+$0.015$\pm$0.006 & 22.530 & $-$0.934  \nl
WF3  & F814W & 22 & $+$0.023$\pm$0.011 & 21.638 & $-$1.818  \nl
WF4  & F555W & 14 & $+$0.009$\pm$0.009 & 22.506 & $-$0.964  \nl
WF4  & F814W & 14 & $-$0.005$\pm$0.011 & 21.609 & $-$1.875  \nl
\enddata
\tablenotetext{^a}{WFPC2 ZP values from Hill et al. (1998, Table 7).\hfill \\}
\tablenotetext{^b}{ALLFRAME instrumental F555W/F814W magnitudes (m$_{\rm inst}$) are obtained 
from the fitted magnitudes (m$_{\rm fit}$) using  the relation: 
${\rm m}_{\rm inst} = {\rm m}_{\rm fit} + 2.5 \times {\rm log}t + {\rm C}$,
where ${\rm C} = -25.00 + 0.016 + 2.5 \times {\rm log}4 + {\rm AC} + {\rm ZP}$,
and $t$ is the exposure time in seconds.
The terms in the C coefficient include: $-$25.00 as the ALLFRAME 
zeropoint, $+$0.016 due to the applied pixel area normalization, a 2.5$\log$4 
term since the data have been converted to integer format and multiplied 
by a factor of 4,  and aperture correction (AC) and WFPC2 zeropoint (ZP) 
terms as given above.}
\end{deluxetable}

\begin{deluxetable}{ccccccccc}
\tablenum{3}
\tablecolumns{9}
\scriptsize
\tablecaption{DoPhot Aperture Correction Coefficients\tablenotemark{a} ~and Zeropoints\tablenotemark{b}}
\tablehead{
\colhead{Filter} &
\colhead{Chip} &
\colhead{CF0} &
\colhead{CF1} &
\colhead{CF2} &
\colhead{CF3} &
\colhead{CF4} &
\colhead{CF5} &
\colhead{ZP} 
}
\startdata
F555W &PC1& $-$0.880 & $-$6.43E$-$05 & $+$1.49E$-$04 & $-$2.00E$-$07 & $-$3.66E$-$07 &$+$5.31E$-$07 & $-$7.677 \nl
F555W &WF2& $-$0.702 & $+$1.67E$-$04 & $+$2.63E$-$04 & $-$2.34E$-$07 & $-$3.02E$-$07 &$+$4.66E$-$08 & $-$7.586 \nl
F555W &WF3& $-$0.614 & $+$1.16E$-$06 & $-$1.02E$-$04 & $-$1.24E$-$07 & $-$4.39E$-$08 &$+$2.67E$-$07 & $-$7.599 \nl
F555W &WF4& $-$0.682 & $+$3.37E$-$04 & $+$1.34E$-$04 & $-$4.68E$-$07 & $-$3.46E$-$07 &$+$2.15E$-$07 & $-$7.601 \nl
      &   &           &               &             &              &              &             &        \nl
F814W &PC1& $+$1.065  & $+$4.33E$-$05 & $+$1.46E$-$04 & $-$2.29E$-$07 & $-$3.11E$-$07 &$+$3.98E$-$07 & $-$8.548 \nl 
F814W &WF2& $-$0.740 & $+$4.31E$-$05 & $+$1.07E$-$04 & $-$1.71E$-$07 & $-$2.32E$-$07 &$+$2.18E$-$07 & $-$8.416 \nl
F814W &WF3& $-$0.763 & $+$4.00E$-$05 & $+$4.78E$-$05 & $-$3.78E$-$08 & $-$6.64E$-$08 &$+$5.68E$-$09 & $-$8.472 \nl 
F814W &WF4& $-$0.764 & $+$1.92E$-$04 & $+$1.27E$-$04 & $-$2.63E$-$07 & $-$2.52E$-$07 &$+$6.88E$-$08 & $-$8.471 \nl
\enddata
\tablenotetext{a}{Aperture correction coefficients determined from observation of Leo I dwarf galaxy field.\\
Aperture correction value: AC = CF0 $+$ CF1$\times$X $+$ CF2$\times$Y $+$ CF3$\times$X$^2$ 
$+$ CF4$\times$Y$^2$ $+$ CF5$\times$X$\times$Y.}
\tablenotetext{b}{DoPhot zeropoints place photometry onto  WFPC2 Ground System of Holtzmann et al. (1995).}
\end{deluxetable}


\begin{deluxetable}{ccccc}
\tablenum{4}
\tablecolumns{5}
\tablewidth{0pc}
\scriptsize
\tablecaption{Comparison of DoPHOT and ALLFRAME Photometry}
\tablehead{
\colhead{Set} &
\colhead{Chip} &
\colhead{N(stars)} &
\colhead{$\Delta {\rm V} \pm \sigma_{\rm V}$} &
\colhead{$\Delta {\rm I} \pm \sigma_{\rm I}$} 
}
\startdata
Secondary  & WF2 & 55 & $-$0.04 (0.01) & $-$0.02 (0.02) \nl
Standards  & WF3 & 62 & $-$0.07 (0.01) & $-$0.07 (0.01) \nl
           & WF4 & 20 & $+$0.04 (0.03) & $+$0.05 (0.01) \nl
\noalign {\smallskip}
\tableline
\noalign {\smallskip}
Cepheids   & WF2 & 10 & $-$0.12 (0.08)  & $+$0.01 (0.15) \nl
           & WF3 &  6 & $-$0.03 (0.06)  & $-$0.11 (0.12) \nl 
           & WF4 &  1 & $+$0.02 (...)   & $+$0.27 (...)  \nl
\enddata
\tablenotetext{}{Note: The magnitude differences listed are in the sense
DoPHOT $-$ ALLFRAME, and the quoted 
errors represent the standard deviation of the mean difference.}
\end{deluxetable}


\begin{deluxetable}{lc}
\tablenum{8}
\tablecolumns{2}
\tablewidth{0pc}
\scriptsize
\tablecaption{Error Budget in the Distance for NGC~1326-A}
\tablehead{
\colhead{Source of Uncertainty} &
\colhead{Error} \nl
\colhead{ } &
\colhead{(mag)}
}
\startdata
{\bf Cepheid Period Luminosity Calibration} & \nl
(a) LMC True Distance Modulus          & $\pm$0.10 \nl
(b) V-band LMC PL Zero Point           & $\pm$0.05 \nl
(c) I-band LMC PL Zero Point           & $\pm$0.03 \nl
(S1) LMC PL Systematic Uncertainty     & $\pm$0.12 \nl
 & \nl
\noalign{\smallskip}
{\bf HST Photometric Calibration}      &  \nl
(d) WFPC2 V-band Photometry Zero Point & $\pm$0.05 \nl
(e) WFPC2 I-band Photometry Zero Point & $\pm$0.05 \nl
(R1) HST Calibration Uncertainty       & $\pm$0.15 \nl
 & \nl
\noalign{\smallskip}
{\bf NGC~1326-A Modulus}               &   \nl
(f) NGC~1326-A V-band PL Fitting        & $\pm$0.06  \nl
(g) NGC~1326-A I-band PL Fitting        & $\pm$0.06  \nl
(R2) Distance Modulus Uncertainty      & $\pm$0.08  \nl
 & \nl
\noalign{\smallskip}
{\bf Metallicity Effect}               &   \nl
(S2) Uncertainty from Metallicity Difference  & $\pm$0.05 \nl
~~~~~ (LMC vs. NGC1326-A)              &   \nl
 & \nl
\noalign{\smallskip}
{\bf Total Uncertainty}                & \nl
(R) Random Errors  (R1$+$R2)           & $\pm$0.17  \nl
(S) Systematic Errors (S1$+$S2)   & $\pm$0.13  \nl
\noalign{\smallskip}
\enddata
\tablenotetext{}{Note: S1 is derived by combining (a), (b), and (c) in
quadrature.  R1 is derived by combining (d) and (e) in quadrature, but
weighted by the reddening coefficients, $R1 = {{[{(1.45)^2(0.05)^2} + 
{(2.45)^2(0.05)^2}]}^{1/2}}$ = 0.15 mag.  
R2 is derived by combining the partially correlated quantities 
(f) and (g) in quadrature.    The individual random and systematic
errors are then combined in quadrature to yield the total random
(R) and systematic (S) errors.}
\end{deluxetable}


\begin{deluxetable}{llllcccc}
\tablenum{13}
\tablecolumns{8}
\tablewidth{0pc}
\scriptsize
\tablecaption{DoPHOT Mean Photometry for Other NGC 1326-A Variables}
\tablehead{
\colhead{ID} &
\colhead{P} &
\colhead{P$_{\rm err}$} &
\colhead{$\chi$$^2$$_r$} &
\colhead{log P} &
\colhead{V$_{ph}$} &
\colhead{I$_{ph}$} &
\colhead{V$-$I$_{ph}$}
}
\startdata
\noalign{\smallskip}
 V1  & $>$60~& ...  &  7.3 &$>$1.78 & 24.65 & 22.73 & 1.92 \nl
 V2  & $>$60~& ...  &  9.0 &$>$1.78 & 26.09 & 24.18 & 1.91 \nl
 V3  &  55~  & (4)  & 12.9 &  1.74 &  24.22 & 23.35 & 0.87 \nl
 V4  &  52.  & (5)  &  7.5 &  1.72 &  22.87 & 21.44 & 1.43 \nl
 V5  & $>$50~& ...  & 17.8 &  1.70 &  23.38 & 22.77 & 0.61 \nl
 V6  & $>$40~& ...  & 11.9 &  1.60 &  24.80 & 23.06 & 1.74 \nl
 V7  &  36.0 & (5)  & 12.4 &  1.56 &  25.96 & 24.77 & 1.19 \nl       
 V8  &  30.5 & (4)  &  4.3 &  1.48 &  26.15 & 25.39 & 0.76 \nl      
 V9  &  25.  & (2)  &  5.4 &  1.40 &  25.89 & 24.86 & 1.03 \nl
 V10 &  22.8 & (1.5)&  8.2 &  1.36 &  25.92 & 25.38 & 0.54 \nl
 V11 &  22.8 & (1.2)&  8.9 &  1.36 &  26.67 & 25.74 & 0.93 \nl
 V12 &  20.7 & (2)  &  4.3 &  1.32 &  25.98 & 24.90 & 1.08 \nl
 V13 &  17.4 & (1)  &  4.5 &  1.24 &  26.34 & 25.23 & 1.11 \nl
 V14 &  16.2 & (1.5)&  2.3 &  1.21 &  25.92 & 25.18 & 0.74 \nl
 V15 &  16.0 & (1)  &  4.5 &  1.20 &  26.79 & ...   & ...  \nl
 V16 &  15.5 & (0.5)&  8.5 &  1.19 &  26.81 & 26.52 & 0.29 \nl
 V17 &  12.6 & (1)  &  2.9 &  1.10 &  27.09 & ...   & ...  \nl
\noalign{\smallskip}
\enddata
\scriptsize
\end{deluxetable}%


\begin{references} 
\reference{} Biretta, J., et al. 1996, Wide Field and Planetary Camera 2 Instrument 
Handbook, version 4.0 (Baltimore: STScI)
\reference{} da Costa, L.N., Pellegrini, P.S., Davis, M., Meiksin, A., Sargent, L.W., 
\& Tonry, J.L. 1991, ApJS, 75, 935 
\reference{} de Vaucouleurs, G., et al. 1991, Third Reference Catalogue of Bright
Galaxies (New York: Springer-Verlag) 
\reference{} Ferguson, H.C. 1989, AJ, 98, 367 
\reference{} Ferrarese, L., et al. 1998, ApJ, 507, 655 (Paper XII)
\reference{} Freedman, W.L., et al. 1994, ApJ, 427, 628 (Paper I)
\reference{} Graham, J.A., et al. 1999, ApJ, in press (Paper XX)
\reference{} Hill, R., et al. 1998, ApJ, 496, 648 (Paper V)
\reference{} Holtzmann, J., et al. 1995a, PASP, 107, 156
\reference{} Holtzmann, J., et al. 1995b, PASP, 107, 1065 
\reference{} Kennicutt, R.C., Freedman, W.L., \& Mould, J.R. 1995, AJ, 110, 1476
\reference{} Kennicutt, R.C., et al. 1998, ApJ, 498, 181  (Paper XIII)
\reference{} Kennicutt, R.C., Huchra, J.P., Macri, L.M., \& Mould, J.R. 1999,
  in preparation
\reference{} Lafler, J., \& Kinman, T.D. 1965, ApJS, 11, 216
\reference{} Madore, B.F., \& Freedman, W.L. 1991, PASP, 103, 933
\reference{} Madore, B.F., et al. 1998, Nature, 395, 47
\reference{} Madore, B.F., et al. 1999, ApJ, 515, 29  (Paper XV)
\reference{} Mould, J.R. et al. 1999, ApJ, submitted (Paper XXI)
\reference{} Phelps, R.L., et al. 1998, ApJ, 500, 763 (Paper IX)
\reference{} Saha, A., \& Hoessel, J.G. 1990, AJ, 99, 97
\reference{} Saha, A., Labhardt, L., Schwengeler, H., Macchetto, F.D., Panagia, N.,
Sandage, A., \& Tammann, G.A. 1994, ApJ, 425, 14
\reference{} Saha, A., Sandage, A., Labhardt, L., Tammann, G.A., Macchetto, 
F.D., \& Panagia, N. 1996, ApJ 46, 55
\reference{} Schechter, P.L., Mateo, M., \& Saha, A. 1993, PASP, 105, 1342
\reference{} Silbermann, N.A., et al. 1999, ApJ, 515, 1 (Paper XIV)
\reference{} Stetson, P.B. 1990, PASP, 102, 932
\reference{} Stetson, P.B. 1994, AJ, 106, 205
\reference{} Stetson, P.B. 1996, PASP, 108, 851
\reference{} Stetson, P.B. 1998, PASP, 110, 1118
\reference{} Stetson, P.B. et al. 1998, ApJ, 508, 491 (Paper XVI)
\reference{} Turner, A. et al., 1998, ApJ, 505, 207 (Paper XI)
\reference{} Welch, D.L., \& Stetson, P.B. 1993, AJ, 105, 1813
\reference{} Zaritsky, D., Kennicutt, R.C., \& Huchra, J.P. 1994, ApJ, 420, 87
\end{references}
\end{document}